\documentstyle[12pt,aaspp4]{article}
\input epsf.sty  
\begin{document}

\title{X-raying the Star Formation History of the Universe} 

\author{A. Cavaliere$^1$, R. Giacconi$^{2}$, and N. Menci$^{3}$}

\affil{$^1$ Astrofisica, Dip. Fisica 2a Universit\`a, Roma I-00133}
\affil{$^2$ 
Johns Hopkins Univ., Dept. of Physics \& Astronomy, Baltimore, 
MD 21218}
\affil{$^3$ Osservatorio Astronomico di Roma, Monteporzio, I-00044}

\baselineskip=22pt

\def\lsim{\ \raise -2.truept\hbox{\rlap{\hbox{$\sim$}}\raise5.truept  
\hbox{$<$}\ }}                                                          
\def\gsim{\ \raise -2.truept\hbox{\rlap{\hbox{$\sim$}}\raise5.truept  
\hbox{$>$}\ }}                                                          
\def\msun{M_{\odot}}

\begin{abstract}

The current models of early star and galaxy formation are based upon the
hierarchical growth of dark matter halos, within which the baryons
condense into stars after cooling down from a hot diffuse phase. The latter 
is replenished by infall of outer gas into the halo potential 
wells; this includes a fraction previously expelled and preheated,  
due to momentum and energy fed back by the SNe which follow the star formation.
We identify such an implied hot phase with the medium 
known to radiate powerful X-rays in clusters 
and in groups of galaxies. We show that the
amount of the hot component required by the current star formation models is
enough to be observable out to redshifts $z \approx 1.5$ in 
forthcoming deep surveys from {\it Chandra} and {\it {\it XMM}}, especially in case the 
star formation rate is high at such and earlier $z$.
These X-ray emissions constitute a necessary counterpart, and will
provide a much wanted probe of the SF process itself (in particular,  
of the SN feedback), to parallel and complement the currently debated 
data from optical and IR observations of the young stars.

\end{abstract}


\section{Introduction}

Is there any link of the early star formation rate (SFR) with extended 
extragalactic sources expected to appear in the deep X-ray surveys 
planned (see Giacconi 1998) with {\it Chandra} and with {\it XMM}?

We expect some such connection to exist, if we carry a step further 
the views started 
by several teams (Munich: Kauffmann, 
White \& Guiderdoni 1993; Durham: Cole et al. 1994, 
Baugh et al. 1998; Santa Cruz: Somerville \& Primack 1998) 
to describe the
processes that lead to galaxy and star formation. These are based upon
the hierarchical growth of structures gravitationally dominated by 
dark matter (DM) halos; from the size of a galaxy to that of a galaxy group
and then of a cluster, the growth 
occurs by repeated merging of smaller into larger structures. 
These views envision the baryons as condensing into stars at the
minima of the forming DM potential wells, upon cooling down from a 
diffuse {\it hot} phase at the virial temperature of the wells.

We will show that such {\it implied} hot component 
-- constituting, in fact, the intracluster medium (ICM) in the larger halos -- 
yields copious bremsstrahlung emissions observable 
in {\it X-rays}. These are closely linked with the optical or infrared 
stellar light 
from which the SFR is currently inferred. The link is 
provided by the energy and momentum fed back 
into the hot phase by the stellar winds and the 
supernovae (SNe) following the star formation. This 
process is especially relevant in the early, small but dense halos; 
some feedback is essential there to 
prevent the runaway cooling of all the contained baryons, 
the so-called  
cooling catastrophe (White \& Rees 1978). 

\section{Key quantities} 

SN explosions, with
some contribution from stellar winds (Bressan, Chiosi \& Fagotto 1994), 
provide an energy output $E_* = $ $E_{SN} \, \eta_{SN}\, \Delta m_*$; here  
$E_{SN} \approx 10^{51}$ erg is the energy of 
a Type II SN explosion, and 
$\eta_{SN} \approx 4\, 10^{-3} \Delta m_*$ 
is the combined efficiency for making SNe and massive blue stars when the mass 
$\Delta m_*$ condenses into the IMF. 

Two key parameters will gauge the
effectiveness of the feedback against the depth of the containing
potential well. Recall that the depth,  
for a halo of mass $M$ with internal density 
$\rho \propto (1+z)^3$ following the background's, 
 is marked by the circular velocity $v = \sqrt {GM/R_v}
\propto M^{1/3}\, \rho^{1/6}$, or by the virial temperature $kT_v \propto 
M/R_v \propto v^2$, which takes on values around 4 keV in rich clusters. 

One parameter will gauge the {\it dynamical} effect of stars and of SN
explosions by way of momentum transfer onto the surrounding gas; 
the resulting galactic winds can directly expel a fraction $\Delta m_h/m_h
\equiv f_* $, especially from shallow potential wells. 
The proper parameter is the fractional energy 
converted into bulk kinetic energy at the escape velocity 
\begin{equation} \epsilon_o = f_* \, m_h \, v^2 /E_{*} ~~.
\end{equation}
This will take on values of order $ 10^{-1}$ in wells with $v \approx
150$ km/s; but equally important will be its differential behavior 
in shallower and deeper wells.

The other parameter $T_v/T_*$ will gauge the importance of the {\it
thermal} effect of the SNe.  Here $k\,T_v$ measures the
gravitational energy of the outer baryons, amenable to conversion into heat
as they infall into the wells.  The ``stellar'' temperature $k T_* =
(1-\epsilon_o) E_{*}\,m_p/3\, \Delta m_h$
measures the preheating level of the expelled fraction, provided by the SNe.
This may be preliminarly estimated as 
\begin{equation}
   k T_* = 0.7 \, (1-\epsilon_o)\,   \Delta m_*/\Delta m_h \gtrsim 0.2~ keV ~~, 
\end{equation}
for a stellar baryonic fraction exceeding $1/5$, the value appropriate only 
for rich clusters (see Renzini
1997). But $kT_v \approx 0.2$ keV is also the virial value corresponding 
to $M\sim 5\, 10^{12}\, M_{\odot}$; so we may surmise that
somewhere in the range from a poor group to a large galaxy 
stellar preheating starts to prevent the gas from piling up 
within the wells. Such masses begin to virialize at $z \approx 2 \pm 0.5$ in
current hierarchical cosmogonies.

We shall investigate whether 
more SF at early $z$, which requires
more baryons condensed within the small halos, also leads to more hot gas  
retained and so to stronger and widespread X-ray emissions. 
To proceed beyond estimates we need a
model for the partition, shifting with $M$ and $z$, of the
baryons among the condensed, the cool, the expelled and the hot
components.

\section{The model} 

\subsection{Optical emissions from stars}
 
To that effect we start from the
semi-analytic models (SAMs) developed by the teams of Munich, Durham 
and Santa Cruz.
These include the ``merging
histories'' of  DM halos as they grow hierarchically through stochastic 
merging events; the above authors describe such histories 
using Monte Carlo simulations. 
We use instead the equivalent analytic probabilities 
provided by Bower (1991) and by Lacey
\& Cole (1993), the so-called extended Press \& Schechter theory. This is 
because our model is considerably more complex as 
it covers also the X-ray emitting baryons; so we need to cut down the
computer time required to average our many observables by 
convolutions over the merging histories.

Along with such a rendition of the DM dynamics, the numerical package devised 
and built by one of us (N. Menci) includes the
current SAM ``recipes'' 
to follow 
the condensed, the cool and the hot baryons. Here we adopt 
the Durham implementation where: the mass of cool gas $\Delta m_c$
is what can cool down at the center of
the halos over their survival time; the SF from the cooling stuff
obeys $\Delta m_* = [m_c - m_* - m_h]\, \Delta t/\tau_*^o\,
(v/300\, km s^{-1})^{\alpha_*}$; the fraction returned from the cool to the hot
component by galactic winds is given by $\Delta m_h$ $= \Delta
m_*(v_h/v)^{\alpha_h}$, corresponding to an effective
$\epsilon_o=\Delta m_h\, v^2/E_* \propto
v^{2-\alpha_h}$. The luminosities of the stars so produced are
computed on convolving $\dot m_*$ with the spectral energy
distributions provided by the model of stellar population synthesis by 
Bruzual \& Charlot (1998).

The primary parameters appearing here are $\tau_{o*}, \alpha_*,
\alpha_h, v_h$; reasonable values for these are known to produce
reasonable first approximations to the galaxy Tully-Fisher relation, 
and to their colors and related luminosity functions. In closer detail, the
latter are improved (as discussed in Cole et 
al. 1994; Somerville \& Primack 1998) on introducing secondary parameters 
to modulate the shape of the bright end (like 
the amount of absorbing dust) and the shape of 
the local, faint end (like the coalescence rate of the galactic baryonic 
cores within the halos). Such galactic parameters are implemented
in our package after the Durham version, but in fact 
do not affect the hot {\it diffuse} 
ICM pervading the halos.  

In this Letter we spare details and figures concerning 
the O and IR observables used here as a calibration, and refer to Menci \& 
Cavaliere (1999; MC99) and Poli et al. (1999). 
We confine ourselves to show in fig. 1 (top) our results concerning 
the intrinsic SFR corresponding to prompt blue 
stellar light. When we use the original Durham 
values for the parameters, 
we obtain optical results equal to theirs, 
see solid line in fig. 1 (top) and 
the related caption.

\subsection{X-ray emissions from the hot phase}

Our {\it new} step is the X-ray emission from the {\it hot}
component (ICM) required by such SFRs. In terms of the ICM temperature $T$ and
particle density $n$, the continuum luminosity reads $L_X
\propto n^2 \, R^3_X\, T^{1/2}$. Using 
$T \approx T_v\propto \rho^{1/3} M^{1/3}$ the scaling is 
conveniently recast as 
\begin{equation}
L_X\propto \rho^{1/2}\;(m_h/M)^2\;(n_2/n_1)^2\;{\cal I}\; T^2 ~.
\end{equation}
The shape factor ${\cal I}$ depends 
only weakly on $z$ and on halo mass $M$; it arises from the 
volume integration over the inner ICM profiles 
$n^2(r)$ and $T^{1/2}(r)$, computed (see CMT99) 
on assuming for the ICM hydrostatic equilibrium in the 
DM potential provided by Navarro, Frenk \& White (1997). 
The main dependences arise from expulsion and heating that 
affect the fraction $m_h/M$ of hot ICM inside the wells, 
and its internal density $n_{2}$ relative to the external $n_{1}$. 
In fact, taking constant $n_2/n_1$ and $m_h/M$ would give 
$L_X \propto T^2$, the scale-invariant 
relation for ICM just following the 
gravitational, hierarchical clustering;  
but the result would clash against the observations 
referred to below. To break such scale-invariance 
the ICM must respond {\it actively}   
 under the drive of the stellar feedback.  

The thermal feedback affects the ratio $n_2/n_1$, which depends 
on the strength of the accretion shocks (see CMT97) induced 
at about the virial 
radius $R_v$ when the outer gas falls supersonically into the halo $M$. 
The latter is built up hierachically by merging with, or 
by accretion of lumps of mass $M'$; the associated 
gas is at a temperature $T'$. 
Given this, the 
density ratio across the shock depends on its strength $T/T'$
after the functional form  $ g (T/T')$ given by  CMT99; this rises steeply 
from 1 when $T/T'$ is moderately increased, 
but saturates to 4 for large values of $T/T'$.

The dynamical feedback modulates this behavior, as it yields two values for 
$T'$: 
the ``stellar'' value $ T_*'$ applies to 
the fraction $f_*= \Delta m_h/m_h$ ejected beyond the virial radius of $M'$ 
and preheated; the gravitational $T_v'$ applies to 
the complementary $1-f_*$ that remains virialized inside the lump. 
Both such components fall into the main well, and the {\it weighted} density ratio 
is given by (MC99) 
\begin{equation}
(n_2/n_1)^2=f_*\,g^2(T/T_*')+(1-f_*)\, g^2(T/T_v')~.
\end{equation}

Meanwhile, some adiabatic compression takes place in the
settling of the shocked ICM into the well. 
Compressions and shocks occur 
repeatedly during the hierarchical growth. Their balance 
may be expressed with a polytropic equilibrium, implying the 
radial profile $T(r) \propto n^{\Gamma -1}(r)$ with $1 < \Gamma
<1.3$ (see Cavaliere \& Fusco-Femiano 1978, Tozzi \& Norman
1999); this converts the
previous history of shocks and compressions in the cluster into a
space stratification. Such ICM profiles are used to compute the 
shape factor ${\cal I}$.

Averaging eq. (4) over all merging histories (i.e., over the 
distributions of $M'$) that lead to $M$  
yields the bent $L_X - T$ relations in fig. 1 (middle); these start steep at 
the scale of groups, gradually flatten, and saturate toward $L_X\propto T^2$ 
only in very rich clusters, in agreement with the observations   
by Ponman et al. (1996); Mushotzky \& Scharf (1997); 
Markevitch (1998); Allen \& Fabian (1998). 
Such a bent shape arises basically because in groups 
the stellar preheating temperature 
$T_*'$ matches or exceeds the virial $T$ as anticipated in Sect. 2; then 
$T/T'$ and $n_2/n_1$ approach 1, so 
$L_X \propto n^2_2/n^2_1$ is weak; in fact, it is weaker when the 
expelled/heated fraction $f_*$ is larger. In rich clusters instead, 
$T$ exceeds any external $T'_*$ and the shocks are always strong; there 
$n_2/n_1$ saturates to 4, and $L_X$ saturates toward $L_X \propto T^2$. 
Additionally, a larger expelled fraction $f_*$ decreases the relative 
amount $m_h/M$ of ICM inside the shallow wells, and again weakens $L_X$ 
in groups.

\section{Results}

All that joins nicely with the 
SAMs, which are based on the same hierarchical merging histories 
as said above. 
In fact, our treatment of the hot ICM can be 
grafted 
onto the SAMs because it proceeds from the same basic equations and 
uses the same  parameters. 
A double bonus of our approach is that heating the gas while expelling 
it outwards to low densities is a process best suited to generate in groups 
the high ``entropy floor'' observed by Ponman Cannon \& Navarro (1999); 
in fact, we obtain $kT_*/n^{2/3} \sim 0.5\, 10^{8/3} \sim 10^2$ 
keV cm$^2$. 
Such a value is conserved near the center 
of forming clusters by the subsequent adiabatic compressions.
But as a cluster forms, increasingly strong shocks develop 
farther out and deposit the entropy 
$\Delta S \propto ln \, T\, n_{1}^{2/3} /T'n_{2}^{2/3}$; 
so the entropy rises outwards,  in accord with 
what is observed in clusters (David, Jones \& Forman 1996; Ponman et al. 1999). 

The derived $L_X-T$ relation constitutes the
intermediate step (analogous to the Tully-Fisher relation for the star light) 
 to compute the X-ray luminosity functions and the
source counts using standard formulae.
We have included the 
line emissions using a routine updating Raymond \& Smith (1977), kindly 
provided by P. Tozzi. 

Our fiducial cosmogony/cosmology 
is provided by standard cold DM initial perturbations in a flat, low-density
Universe with $\Omega_o=0.3$ and $\Omega_{\Lambda}=0.7$, and $H_0$=70
km/s Mpc.  Not only this is indicated by the data concerning 
SNe Ia, but also it makes the baryonic fraction $\Omega_b=0.04$ 
suggested by the cosmological nucleosynthesis (see Olive 1998)
consistent with the values 
$\Omega_b/\Omega_0$ around 0.15 from mass inventories in clusters
(White et al. 1993; Ettori \& Fabian 1999). 

So equipped, we explore the counterparts in X-rays of two extreme
evolutions of the SFR for $z>1$ (shown in fig. 1, top), which bracket the 
currently debated
shapes derived from the O-IR data. 
The first SFR {\it declining} for $z\gtrsim 1.5$ is 
obtained from the original Durham parameters 
(see specifically Heyl et al. 1995). 
Their values $\alpha_*=1.5$ and 
$\alpha_h=5.5$ {\it minimize} the star formation and 
{\it maximize} the feedback effects in the shallow, early wells;  
in particular, the value of $\alpha_h $ corresponds to 
 $\epsilon_o = v^{-3}$, that is, to galactic winds 
{\it stronger} in shallower wells. The corresponding 
$L_X-T$ relation at $z = 0,\, 1$ 
is shown in the middle panel of fig. 1 (left), 
and the prediction for the soft X-ray 
counts by the solid line in the bottom panel of fig. 1.

The other extreme evolution is the {\it flat} SFR, also shown in
fig. 1 (top); this is derived from the second set of parameter values listed in the 
caption. The set still leads to comparable optical 
luminosity functions but is more in tune with recent 
feedback reappraisals toward the low side (see Thornton et al. 1998, 
Ferrara \& Tolstoy 1999; Martin 1999); it includes the ``neutral'' values 
$\alpha_*=0$ and $\alpha_h = 2$. 
These correspond to the SF time $\tau_* \propto 
v^{-\alpha_*}$ and the kinetic fraction $\epsilon_o \propto 
v^{2-\alpha_h}$ being {\it even} for shallow and deep wells; 
in particular, the values of $v_h, \alpha_h$ correspond
to $\epsilon_o=\, $const$\sim 10^{-1}$ in the shallow wells, that is, 
to {\it weaker} winds. In the middle panel 
of fig. 1 (right) we show the corresponding $L_X-T$ relation, and 
by the dashed line in the bottom panel of fig. 1 the corresponding counts.

We stress that the deep counts in fig. 1 (bottom) 
draw increasingly {\it apart} for fluxes $F_X <   10^{-14}$ erg
cm$^{-2}$ s$^{-1}$; in fact, at fluxes 10 times fainter the counts
for weak  
exceed  those for strong winds 
by a factor at least 3.
Following up the discussion ending \S 3.2, 
the result can be reckoned on the basis of 
$L_X \propto m^2_h\, n^2_2/ n^2_1$ (see
eqs. 3 and 4). Here $m_h$ $\propto f^{-1}_*$; in addition, $n^2_2/n^2_1$ is
enhanced at low $T$ by about $f^{-1}_*$ for weak compared to strong
feedback; so the luminosities are enhanced as $L_X \propto f^{-3}_*$ at low
$T$. The numerical values of $f_*$ 
 differ by $-20\%$ at $L_X \sim 10^{43}$ erg/s and $z 
\approx 
1$. In sum, with weak feedback we expect the deep counts, which rise here
close to $N(>F_X)\propto F_X^{-3/2}$, to exceed the other case by about
$(0.8)^{- 9/2}\approx 3$ at $F_X \sim 10^{-15}$ erg cm$^{-2}$s $^{-1}$,
as confirmed by fig. 1 (bottom).

\section{Discussion and conclusions}

Having in mind surveys from {\it Chandra} and {\it XMM} we have 
limited  our computations to photon energies exceeeding 
$E_{min}$ $= 0.25$ keV, and 
the resolvable sources to $L_X \geq 10^{43}$ erg/s with 
$F_X \geq 10^{-15}$ erg/cm$^2$s. So we obtain a {\it lower}
bound to the divergence of the counts; this is amplified by
some $20 \%$ if luminosities down to $L_X \sim 3\, 10^{42}$
erg/s are included. In 
the critical universe, a given $L_X$ would 
correspond to a brighter flux, to yield even larger excess 
counts at given flux.

Considerable uncertainties currently plague
the evolution of the SFR to $z > 1$, as derived from the
optical (dust-obscured) and from the IR data (time-consuming to 
obtain, and with redshifts difficult to determine); 
these are discussed in detail, e.g., by Madau, Pozzetti \& Dickinson (1998), 
Pettini et al. (1998), Hughes et al. (1998), Barger et al. (1998), 
Ellis (1998). 
So complementary information concerning the X-ray
counterparts from the forthcoming surveys planned with {\it Chandra} and {\it XMM}
will be welcome and timely to address the issue. Eventually, with the
O-IR data consolidating in their own right, the soft X-rays will 
elicit the SN feedback in action. In fact, this contributes {\it
directly} to the hot component, while it only indirectly affects the
stars which form from the cooling stuff; therefore the soft X-rays are best 
suited to probe this key quantity. 

In fig. 2 we show not only the evolution of the stellar and the hot 
components from our computations, but also the abundances
at various $z$ of the third phase constituted by the {\it lukewarm}
baryons inside and outside the galaxies. These turn out to be
in fair agreement with the densities 
(actually lower bounds) evaluated 
from the observations of the Ly$\alpha$ clouds seen in
absorbtion, whether damped or not. 

In conclusion, a weaker feedback $\Delta m_h/\Delta m_*$ allows 
 more baryons (hot and cold) to be retained within shallow early wells,  
as illustrated in fig. 2. As shown in fig. 1, such baryons yield not only 
a larger star formation but also stronger X-ray emissions.
These involve  mainly outputs 
L$_X \sim 10^{43}$ erg/s or less at $z \approx 1$ 
or larger, associated with halo masses 
$M \sim 5 \, 10^{13}\, M_{\odot}$ or smaller and with 
temperatures $T \approx 1$ keV or cooler.  
The deep counts in soft X-rays of souch sources 
and their $L_X - T$ relation 
can probe {\it directly} the feedback, the main unknown that hampers 
the understanding the cosmic star formation history.

\smallskip
We thank E. Giallongo, P. Rosati, and P. Tozzi 
for stimulating and informative discussions, and a referee 
for helpful comments. Work supported by grants from ASI and MURST.

\newpage
\section*{FIGURE CAPTIONS}

\figcaption[]{
Top: The $z$-dependent SFR from our model of
baryon processing during the hierarchical clustering in the flat 
universe with $\Omega_o = 0.3, H_o= 70$ km/s Mpc. The solid line
represents the shape peaked at $z\approx 1.5$ 
resulting from the Durham set of parameters: $\tau_*^o=2$
Gyr, $\alpha_*=-1.5$, $v_h=140$ km/s, $\alpha_h=5.5$, implying strong galactic 
winds. 
The dashed line represents the SFR flat toward high $z$ resulting from our 
``neutral'' set: 
$\tau_*^0=2$ Gyr, $\alpha_*=0$, $v_h=140$ km/s, $\alpha_{h}=1.5$, implying 
weak winds. 
Middle: The $L_X-T$ correlations at $z=0$ (solid line) and
at $z=1$ (dotted line) corresponding to the peaked (left frame) and
to the flat SFR (right frame) illustrated on top. Group data from
Ponman et al. (1996, solid squares); cluster data from Markevitch
(1998, open circles). 
Bottom: The source counts $N(>F_X)$ corresponding to the 
peaked SFR (solid line) and
to the flat SFR (dashed line), in the energy range 0.25-2 keV.  The
hatched band represents the cluster number counts
observed by Rosati et al. (1998).
\label{fig1}}

\figcaption[]{ The contribution to the baryon density $\Omega_{b}$ 
from the lukewarm baryons with $T \leq 5\, 10^4$ K, from  the hot virialized 
ICM and from the stars, see labels in the figure. 
The solid lines show the model
predictions corresponding to strong winds, while the dashed lines 
correspond to weak winds. 
The density of lukewarm baryons is compared 
with the data (from Giallongo et al. 1997) concerning the density of
photoionized intergalactic gas, which constitute lower limits especially for 
$z<3$.  
\label{fig2}}


\begin{references} 

\reference{}Allen, S.M, \& Fabian, A.C. 1998, MNRAS, 297, L57

\reference{}Barger, A., Cowie, L.L., Sanders, D.B., \& Fulton, E. 1998, 
 Nature, 394, 248


\reference{}Baugh, C.M., Cole, S., Frenk, C.S., \& C.G. Lacey 1998, ApJ, 
 498, 504 

\reference{}Bower, R. 1991, MNRAS, 248, 332

\reference{}Bressan, A, Chiosi, C, \& Fagotto, F. 1994, ApJ SS, 94, 63 

\reference{}Bruzual, A., \& Charlot, S. 1998, update of ----- 1993, 
ApJ, 405, 538

\reference{}Cavaliere, A, \& Fusco Femiano, R. 1978, \aap, 70, 767

\reference{}Cavaliere, A., Menci, N., Tozzi, P., 1997, ApJL 484, 1 (CMT97) 


\reference{}----------------------------------1999, preprint [astro-ph9810498] 

\reference{}Cole, S., Aragon-Salamanca, A., Frenk, C.S., Navarro, J.F., \& 
 Zepf, S.E. 1994, MNRAS, 271, 781

\reference{}David, L.P., Jones, C., \& Forman, W. 1996, ApJ, 473, 692

\reference{}Ettori, S., \& Fabian, A.C. 1999, MNRAS, 305, 843

\reference{}Ellis, R. 1998, Nature, 395, 3

\reference{}Ferrara, A., \& Tolstoy, R. 1999, preprint [astro-ph/9905280]
 
\reference{}Giallongo, E., Fontana, A., \& Madau, P. 1997, MNRAS, 289, 629

\reference{}Giacconi, R. 1998, Astron. Nachr. 319, 147

\reference{}Kauffmann, G., White, S.D.M., \& Guiderdoni, B. 1993, MNRAS, 
  264, 201

\reference{}Heyl, J.S., Cole, S., Frenk, C.S., \& Navarro, J.F. 1995, MNRAS, 
274, 755

\reference{}Hughes, D. et al. 1998, Nature, 394, 241 ]

\reference{}Lacey, C., \& Cole, S. 1993, \mnras, 262, 627

\reference{}Madau, P., Pozzetti, L., \& Dickinson, M.E.
 1998, ApJ, 498, 106

\reference{}Markevitch, M. 1998, ApJ, 504, 27

\reference{}Martin, C. 1999, ApJ, 513, 156

\reference{}Menci, N., \& Cavaliere, A. 1999, MNRAS, in press 
[astro-ph/9909259]

\reference{}Mushotzky, R.F., \& Scharf, C.A., 1997, ApJ, 482, 13

\reference{}Navarro, J.F., Frenk, C.S., \& White, S.D.M. 1997, \apj,
490, 493

\reference{}Olive, K.A. 1998, in Proc. of 19th Texas Symposium on 
Relativistic Astrophysics and Cosmology, eds. J. Paul et al. (Saclay: CEA) 

\reference{}Pettini, M., Kellog, M., Steidel, C.C., Dickinson, M., Adelberger, 
K.L., Giavalisco, M. 1998, ApJ, 508, 539

\reference{}Ponman, T.J., Bourner, P.D.J., Ebeling, H., B\"ohringer, 
H. 1996, MNRAS, 283, 690

\reference{}Ponman, T.J., Cannon, D.B., \& Navarro, J.F. 1999, Nature, 397, 135

\reference{}Poli, F., Giallongo, E., Menci, N., D'Odorico, S., \& Fontana, A. 
1999, ApJ, in press [astro-ph/9910537]

\reference{} Raymond, J.C., \& Smith, B.W. 1977, ApJS, 35, 419

\reference {} Renzini, A. 1997, ApJ 488, 35

\reference{} Rosati, P., Della Ceca, R., Norman, C., \& Giacconi, R. 
1998, ApJ, 492, 21

\reference{}Somerville, R.S., \& Primack, J.R. 1998, 
preprint [astro-ph/9802269]  

\reference{}Tozzi, P. \& Norman, C. 1999, preprint [astro-ph9905046]

\reference{}Thornton, K., Gaudlitz, M., Janka, H.-Th., Steinmetz, M. 1998, 
ApJ, 500, 95

\reference{}White, S.D.M., Navarro, J.F., Frenk, C.S., \& Evrard,
A.E. 1993, Nature, 366, 429

\reference{}White, S.D.M., \& Rees, M.J. 1978, MNRAS, 183, 341


\end{references}
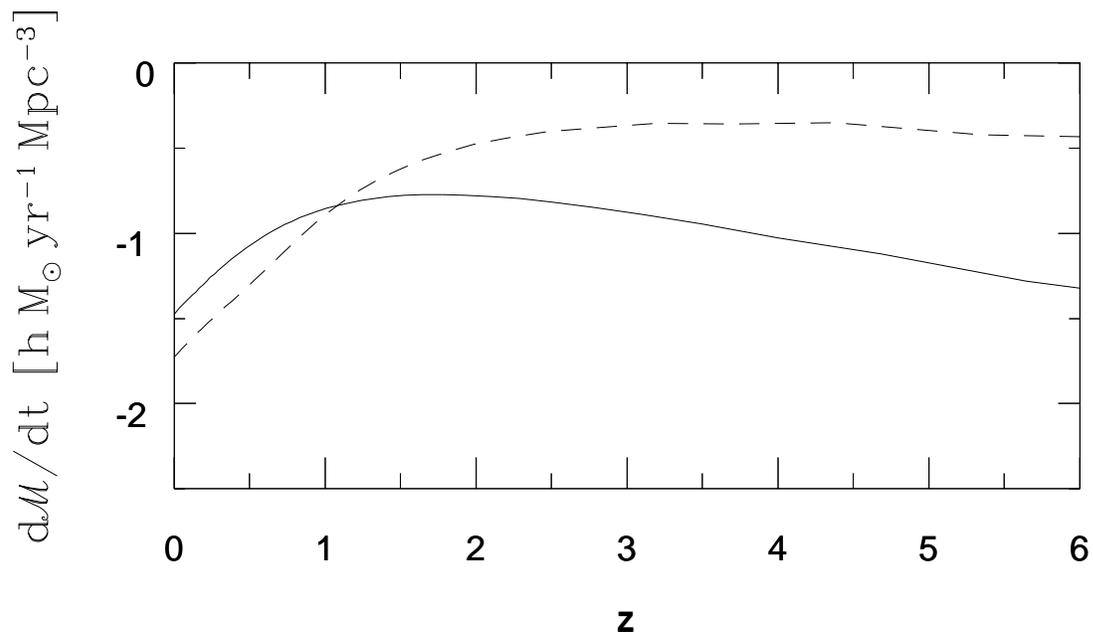
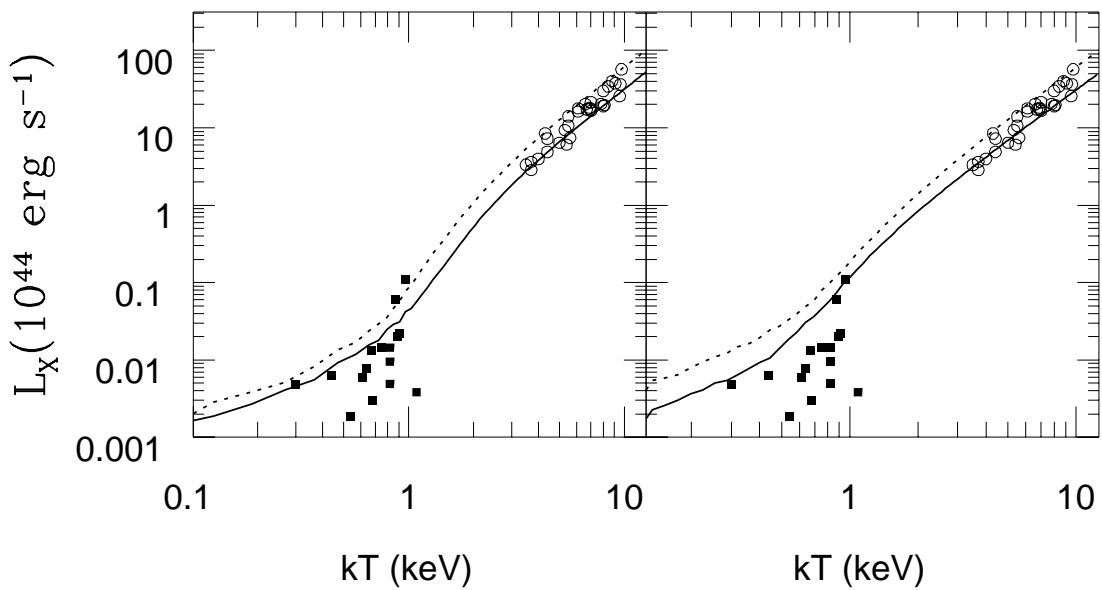
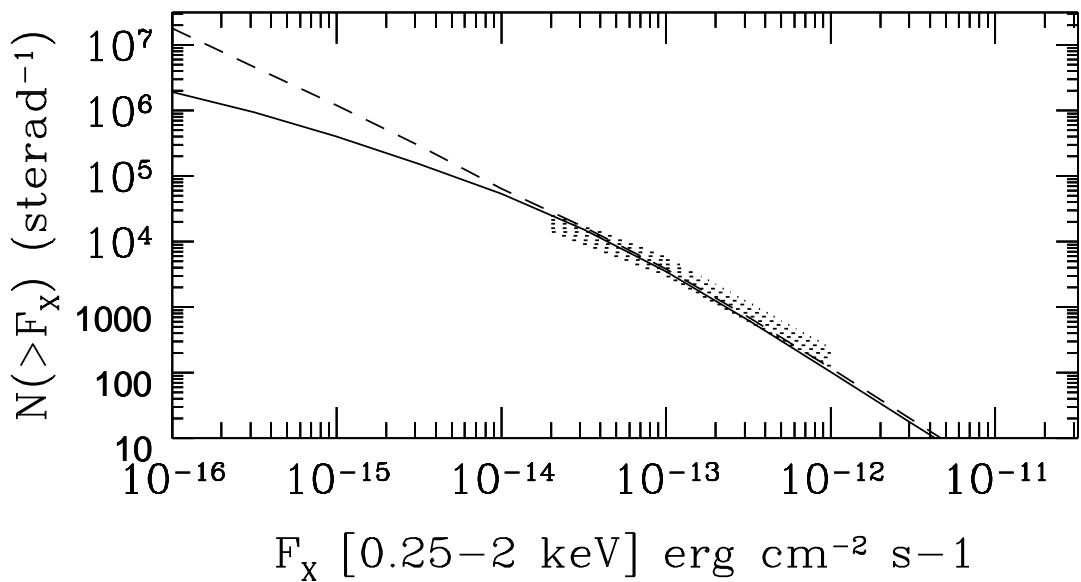
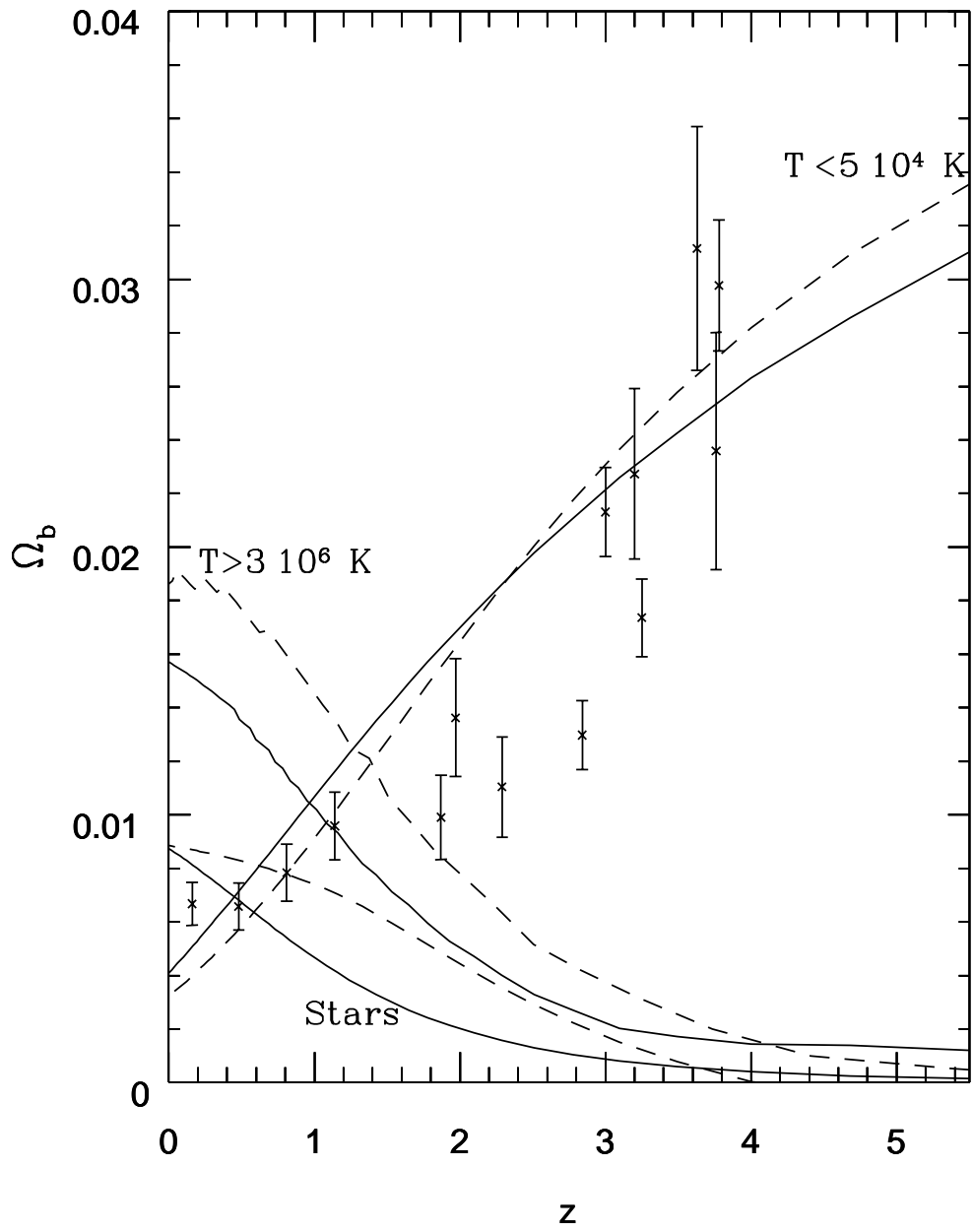
\end{document}